\documentclass{article}
\usepackage{ijcai22}

\usepackage{times}

\usepackage{soul}
\usepackage{url}
\usepackage[hidelinks]{hyperref}
\usepackage[utf8]{inputenc}
\usepackage[small]{caption}
\usepackage{graphicx}
\usepackage{amsmath}
\usepackage{booktabs}
\newenvironment{blockquote}{%
  \par%
  \medskip
  \leftskip=1.5em\rightskip=1.5em%
  \noindent\ignorespaces}{%
  \par\medskip}
  
\title{A New Era: Intelligent Tutoring Systems \\ Will Transform Online Learning for Millions}

\author{
Francois St-Hilaire$^{1}$\and
Dung Do Vu$^{1}$ \and
Antoine Frau$^{1}$ \and
Nathan Burns$^{1}$ \and
Farid Faraji$^{1}$ \and
Joseph Potochny$^{1}$ \and
Stephane Robert$^{1}$ \and 
Arnaud Roussel$^{1}$ \and
Selene Zheng$^{1}$ \and
Taylor Glazier$^{1}$ \and
Junfel Vincent Romano$^{1}$ \and
Robert Belfer$^{1}$ \and
Muhammad Shayan$^{1}$ \and
Ariella Smofsky$^{1}$ \and
Tommy Delarosbil$^{1}$ \and
Seulmin Ahn$^{1}$ \and
Simon Eden-Walker$^{1}$ \and
Kritika Sony$^{1}$ \and
Ansona Onyi Ching$^{1}$ \and
Sabina Elkins$^{1}$ \and
Anush Stepanyan$^{1}$ \and
Adela Matajova$^{1}$ \and
Victor Chen$^{1}$ \and
Hossein Sahraei$^{1}$ \and
Robert Larson$^{1}$ \and
Nadia Markova$^{1}$ \and
Andrew Barkett$^{1}$ \and
Laurent Charlin$^{1, 2}$ \and
Yoshua Bengio$^{1, 2}$ \and \\
Iulian Vlad Serban$^{1}$ \and
Ekaterina Kochmar$^{1, 3}$\\ 
\vspace{0.5em}\affiliations
$^1$Korbit Technologies Inc., Canada\\
$^2$Quebec Artificial Intelligence Institute (Mila), Canada\\
$^3$University of Bath, United Kingdom\\
}

\begin{document}

\maketitle

\begin{abstract}
Despite artificial intelligence (AI) having transformed major aspects of our society, less than a fraction of its potential has been explored, let alone deployed, for education.
AI-powered learning can provide millions of learners with a highly personalized, active and practical learning experience, which is key to successful learning.
This is especially relevant in the context of online learning platforms. 
In this paper, we present the results of a comparative head-to-head study on learning outcomes for two popular online learning platforms (n=199 participants): A {\tt MOOC} platform following a traditional model delivering content using lecture videos and multiple-choice quizzes, and the {\tt Korbit} learning platform providing a highly personalized, active and practical learning experience. 
We observe a huge and statistically significant increase in the learning outcomes, with students on the {\tt Korbit} platform providing full feedback resulting in higher course completion rates and achieving learning gains $2$ to $2.5$ times higher than both students on the {\tt MOOC} platform and students in a control group who don't receive personalized feedback on the {\tt Korbit} platform.
The results demonstrate the tremendous impact that can be achieved with a personalized, active learning AI-powered system.
Making this technology and learning experience available to millions of learners around the world will represent a significant leap forward towards the democratization of education.

\end{abstract}

\section{Introduction}



In the recent years, the field of Artificial Intelligence (AI) has seen many impressive breakthroughs. With the development of large-scale, accurate, and reliable AI systems, society is moving in the direction of AI-system integration and AI-augmentation in many spheres of life.
Of particular importance among those are the areas where AI can improve the living conditions of humans and can benefit the society on the whole, i.e., where {\em AI can be used for global good}. Such applications can include, among others, the use of AI to fight poverty and hunger, address inequality and climate change, facilitate economic growth and innovation, contribute to good health and well-being, and so on. One urgent area to which AI could and should contribute is {\em education}, in particular, ensuring accessibility of high-quality education for people around the globe. In the educational domain, AI-enabled solutions can help ensure inclusive and equitable high-quality education and promote life-long learning opportunities for all in accordance with the United Nations' Sustainable Development Goals (SDGs).\footnote{https://sdgs.un.org/goals/goal4}

Unfortunately, high-quality education is not accessible to the vast majority of people around the world. There are many areas in the world where even basic school infrastructure is missing. This creates problems for in-person learning process, which have been further exacerbated by the ongoing COVID-19 pandemic~\cite{adedoyin2020covid,basilaia2020transition,onyema2020impact}.
Nevertheless, while in-person education may not be available to everyone, the growing availability of the internet and devices such as laptops and smartphones create a unique and critical opportunity to bridge the gap between those who can get access to high-quality in-person teaching and those who can't.

One of the solutions proposed are online learning platforms providing large numbers of students with access to learning material on various subjects, with Massive Online Open Courses (MOOCs) being a notable example~\cite{wang2014longitudinal,tomkins2016predicting}.
Such platforms have the capability to address inequalities in society caused by the uneven access to in-person teaching. However, despite this, it is unclear whether these capabilities have been exploited to the full extent.
There is evidence that MOOCs mostly benefit students who already have degrees and live in developed countries and, thus, may not yet be delivering on their promise to truly ``democratize" education~\cite{wildavsky2015moocs}.
Furthermore, student dropout rates in MOOCs often exceed $90\%$~\cite{rieber2017participation}, with poor interaction between the system and its users identified as being a major cause~\cite{hone2016exploring}. 
Personalization is key to successful and effective learning~\cite{bloom19842}.
Given that MOOCs lack personalization and adaptivity, this can be identified as one of the major reasons for MOOCs not being able to provide high-quality effective education to everyone.
A more powerful solution among computer-based learning environments (CBLEs), which can provide students with a scalable, high-quality alternative is AI-powered Intelligent Tutoring Systems (ITS)~\cite{psotka1988intelligent,graesser2001intelligent,koedinger2006cognitive}.

In this paper, we evaluate online learning and put the assumptions around their learning efficacy to the test.
We conduct a comparative head-to-head study on learning outcomes for two popular online learning platforms. 
Among these, the {\tt MOOC} platform follows a traditional model delivering content over a series of lecture videos and multiple-choice quizzes, while the {\tt Korbit} platform\footnote{{\tt Korbit} is available at \url{www.korbit.ai}.} provides a highly personalized, active learning experience involving problem-solving exercises and personalized feedback.
Learning outcomes are measured on the basis of pre- and post-assessment quizzes with participants taking courses on an introductory data science topic on the two platforms.
As a result, we observe a statistically significant increase in the learning outcomes, with students on {\tt Korbit} providing full feedback achieving learning gains $2$-$2.5$ times higher than both students on the {\tt MOOC} platform and a control group of students who don't receive personalized feedback on the {\tt Korbit} platform.
In addition, we find that students learning on the highly personalized, active learning {\tt Korbit} platform achieve higher course completion rates and are more motivated than all other groups of students.
The major contribution of this paper is the demonstration that personalized, active learning AI-powered systems based on problem-solving exercises and personalized feedback of the type deployed on {\tt Korbit} can have a tremendous impact on the learning experience and substantially improve learning efficacy.
If such a personalized and active learning experience can be made available to millions of learners around the world, then this advance would represent a significant step towards the democratization of education.


\section{Background}

\subsection{Personalization in Computer-based Learning Environments}

In his seminal work, Bloom~\cite{bloom19842} demonstrated that personalised one-on-one tutoring results in learning gains as high as 2 standard deviations ($2\sigma$), which means that the average tutored student scores better than $98\%$ of the students that do not receive one-on-one tutoring. Clearly, one-on-one tutoring is too costly for most societies to provide their citizens with, so in practice this target is not achieved. A viable low-cost alternative to in-person one-on-one tutoring is Intelligent Tutoring Systems (ITS), which are {\em ``computer-based instructional systems with models of instructional content that specify what to teach, and teaching strategies that specify how to teach"}~\cite{Wenger}. Such systems attempt to mimic personalized human tutoring in a computer-based environment~\cite{anderson1985intelligent,nye2014autotutor}.

ITS have been shown to be able to address such personalization factors as individual student characteristics~\cite{Graesser} and cognitive processes~\cite{Wu}, provide personalized feedback and interactions, or even develop personalized curriculum and generate personalized feedback~\cite{chi2011instructional,Lin,rus2014macro,rus2014deeptutor,piano2018,mu2018combining,albacete2019impact,munshi2019personalization,luccioni2016sti,hollander2022intelligent,alshaikh2021ai}. Moreover, previous research suggests that in some domains ITS may have already surpassed non-expert human tutors and matched expert ones~\cite{vanlehn2011relative,graesser2012intelligent}.

\subsection{Korbit Personalized Learning Platform}
\label{ssec:korbit}


In this paper, we investigate the impact of personalization on the quality of education in the context of AI-powered learning. We experiment with the {\tt Korbit} learning platform, which is an AI-powered, large-scale, dialogue-based ITS.
Korbit has trained over $20,000$ learners online with highly personalized, active and practical learning experience across a range of subjects and skills.
The platform uses a fully-automated system based on a suite of machine learning (ML), natural language processing (NLP), and reinforcement learning (RL) models aimed to provide deeply personalized and interactive learning online.
The platform currently hosts a number of courses and skills around software development, data science, machine learning, and artificial intelligence.
It is highly modular, scalable, and is capable of customizing the curriculum for each individual student and adapting in real-time the content to the students' levels of proficiency and needs.
Moreover, the ML, NLP, and RL models at the core of {\tt Korbit} continue to learn online in real-time based on live interactions with students, automatically adjusting themselves to new content and learning activities~\cite{serban2020large}.

To achieve deep personalization and interactivity, {\tt Korbit} alternates between teaching via video lectures, Socratic tutoring, interactive problem-solving exercises, coding exercises and project-based learning.

\begin{figure*}[h!]
\centering
\includegraphics[width=0.9\textwidth]{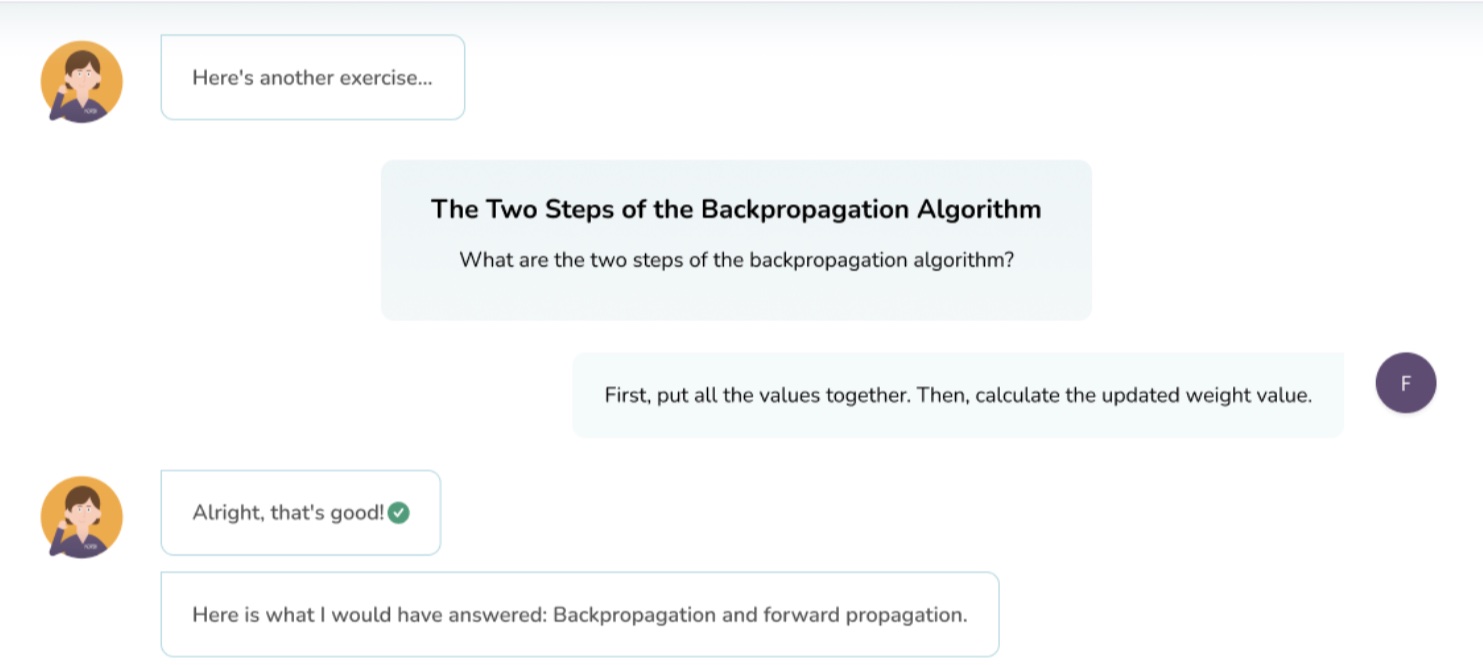}
\caption{An example of an interaction on the platform between {\tt Korbi}, the AI tutor, and a student.} \label{fig:dialogue}
\end{figure*}

The focus of this study is on the problem-solving sessions. Here students are presented with problem statements (e.g.,\@ questions), whereupon the student can attempt to solve the exercise, ask for help, or skip the exercise. If the student attempts to solve the exercise, their solution attempt is compared by an NLP-driven solution verification module against the expectation stored internally in the database (i.e. reference solution, which typically consists of one or two sentences containing all relevant information that should be included in the correct answer to the question posed). Figure \ref{fig:dialogue} presents an example of an interactive dialogue between {\tt Korbi}, the AI tutor, and a student.

In a full interactive, personalized model, if a student's solution is incorrect, the system responds with one of a dozen different pedagogical interventions to help students arrive at the correct solution to the problem. Such pedagogical interventions on the {\tt Korbit} platform include, among others, hints, explanations, elaborations, mathematical hints, concept tree diagrams, and multiple choice quiz answers.
The type and the levels of difficulty for each pedagogical intervention is chosen by RL models based on the student's learning profile and previous solution attempts.
This helps ensure that the content and the learning experience are highly personalized and adapted to each particular student.
In addition to questions, reference solutions, and pedagogical interventions created manually by curriculum designers for the {\tt Korbit} platform, many questions and hints are further automatically generated using ML and NLP models~\cite{grenander2021deep,kochmar2021automated,kulshreshtha2021back}.
The {\tt Korbit} platform is highly scalable and has the ability to parse vasts amounts of open educational resources (OER), automatically generalize to new subjects and improve in real-time as it interacts with new students.

The {\tt Korbit} platform applies a number of personalization strategies: once a learner is registered on the platform, they are presented with a short series of knowledge and skills assessments aimed at identifying their level of expertise and providing them with the most suitable personalized learning path.

\begin{figure*}[h]
\centering
\includegraphics[width=0.9\textwidth]{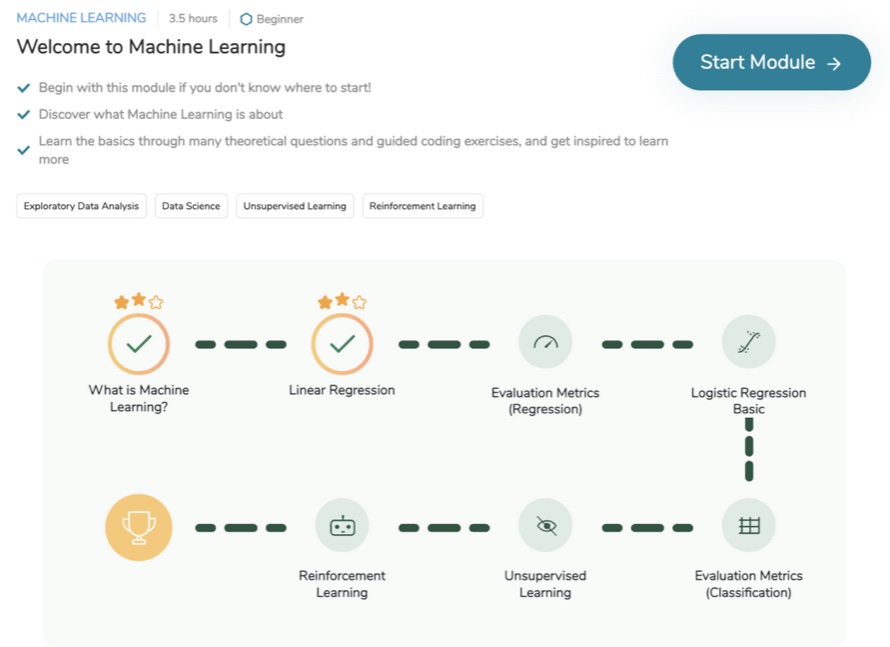}
\caption{An example of a personalized learning path.} \label{fig:curriculum}
\end{figure*}

Once the level of the learner is identified, {\tt Korbi}, the AI tutor, tailors the learning materials and the path based on the learner's understanding of key concepts, time availability, and learning objectives.
The level of difficulty and the learning pace are constantly adapted to the learner's needs based on their performance throughout the learning path.
Figure \ref{fig:curriculum} provides an example of a personalized learning path.

\begin{figure*}[h]
\centering
\includegraphics[width=0.9\textwidth]{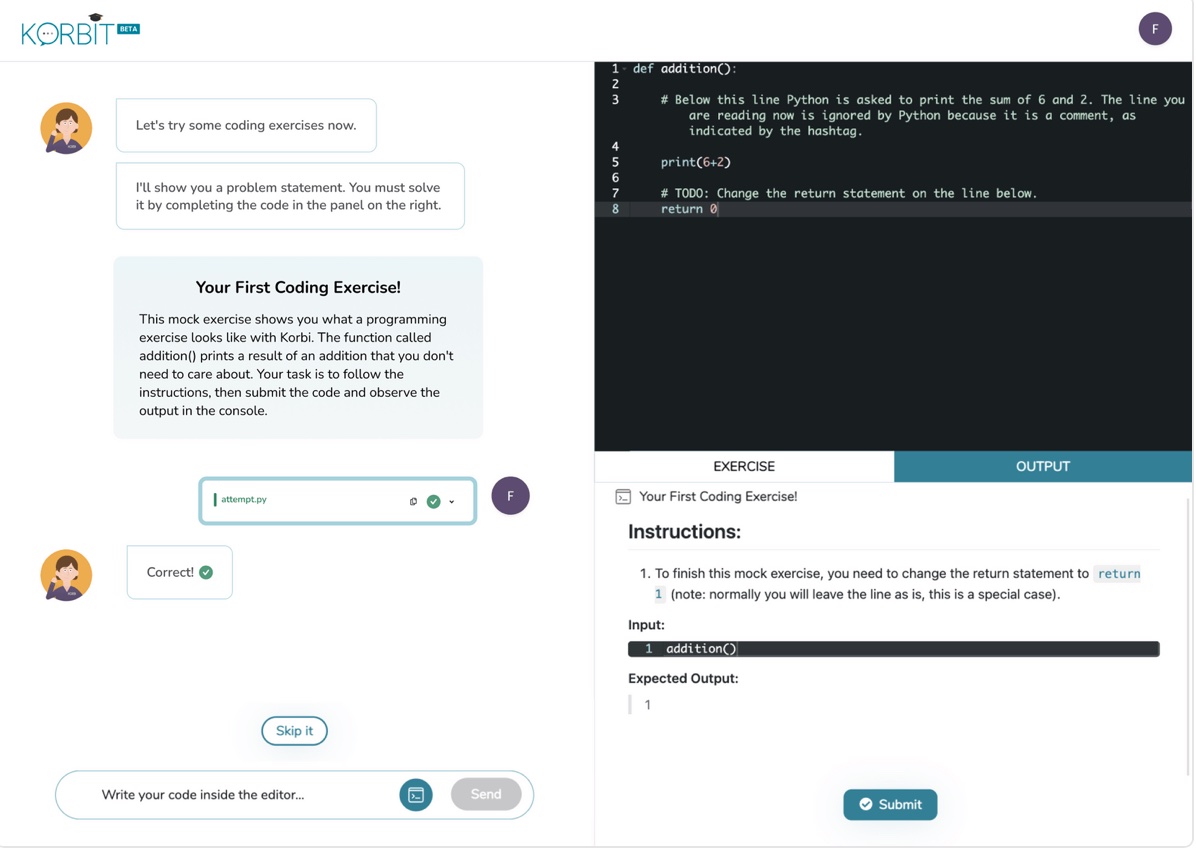}
\caption{An example of a coding exercise on the {\tt Korbit} platform.} \label{fig:coding}
\end{figure*}

In addition to the personalization strategies, {\tt Korbi} promotes active and skill-based learning. Questions, exercises and projects on the platform encourage learners to constantly apply their knowledge in practice.
The platform offers coding exercises and programming projects that allow learners to translate their declarative knowledge into procedural knowledge (see Figure \ref{fig:coding} for an example of a coding exercise).
A recent user study demonstrated that such personalized, active approach to learning leads to significantly improved learning gains as well as higher meta-cognition in learners studying with {\tt Korbit}~\cite{st2021comparative}.

\section{Experimental Setup}
\label{sec:setup}

In this study, we aim to investigate the effect that personalization has on the quality of education provided by online learning systems. This section details our experimental setup.

\subsection{Participants}

This study was run in collaboration with an industrial partner: specifically, to investigate the impact of personalization provided by AI in online and distance learning, we have partnered with an information technology company in Vietnam. Employees of this company have strong programming skills but needed upskilling with respect to their data science and machine learning knowledge and practical skills. 

Over $200$ software developers from the company were originally offered an opportunity to participate in the study and were asked to fill in a short enrolment questionnaire, specifying their level of expertise in data science. As a result, employees with no background knowledge of data science were selected and provided access to a $3$-hour long course on {\em linear regression} on one of the two online platforms. Prior to starting the course, the participants were asked to do a pre-assessment quiz. Then, upon completion of the course on the correspondent platform, participants were asked to do a post-assessment quiz, and their learning outcomes were measured based on the difference in their scores.
The experiment was run completely online. 
Participants who finished either course were rewarded with a completion certificate awarded jointly by us and the employer regardless of their scores, which provided them with an incentive at the company level to participate in the study.

The analysis of the results was performed on the set of participants who qualified for the study, completed it and submitted answers to both pre- and post-assessment quizzes. As a result, we analyzed the submissions from $50$ study participants.

\subsection{Learning Platforms}

Participants, who qualified for the study based on their (lack of) prior experience with data science, were randomly split between two platforms -- a {\tt MOOC} platform and the {\tt Korbit} learning platform. In addition, to highlight the impact of feedback personalization and remove any other factors that may stem from the differences between the two platforms that are not directly related to personalization, we ran experiments on {\tt Korbit} under two settings -- the full interactive personalization mode, and a more limited interactive mode with no personal feedback provided to the students by the AI-tutor. Finally, we also set a threshold of $25\%$ skipped exercises, meaning that a participant who skipped more than $25\%$ of the exercises on the {\tt Korbit} platform would not be considered in the analysis. As a result, we ended up with the sample size of $50$ participants in total, split between the following three groups: 

\begin{itemize}
    \item {\tt MOOC} is a widely popular learning platform with millions of learners that follows a traditional model for online and distance-learning courses: students on this platform learn by watching lecture videos, reading, and testing their knowledge with multiple choice quizzes. As such, this platform does not provide any personalization to students under any settings. This groups consisted of $26$ participants.
    \item {\tt Korbit (full)} is an AI-powered learning platform, which relies on machine learning models to adapt the learning process to students and their performance in real time and to provide them with personalized distance-learning education, as detailed in Section \ref{ssec:korbit}. The group of students allocated to this group got access to the full-functioning system with a pre-defined curriculum matching the {\tt MOOC} course. This group consisted of $16$ participants.
    \item {\tt Korbit (no feedback)}: this group got access to a special variant of {\tt Korbit} with a pre-defined curriculum matching that of the {\tt MOOC} course, but in this case, {\tt Korbit} was not providing any feedback except saying if students' answers were ``CORRECT" or ``INCORRECT". Although we realize that students in this group were more frustrated overall as a result of this limitation, this was an important comparison point which helped us evaluate the impact of the personalized feedback. This group consisted of $8$ participants.
\end{itemize}

\begin{figure*}[t!]
\centering
\includegraphics[width=0.9\textwidth]{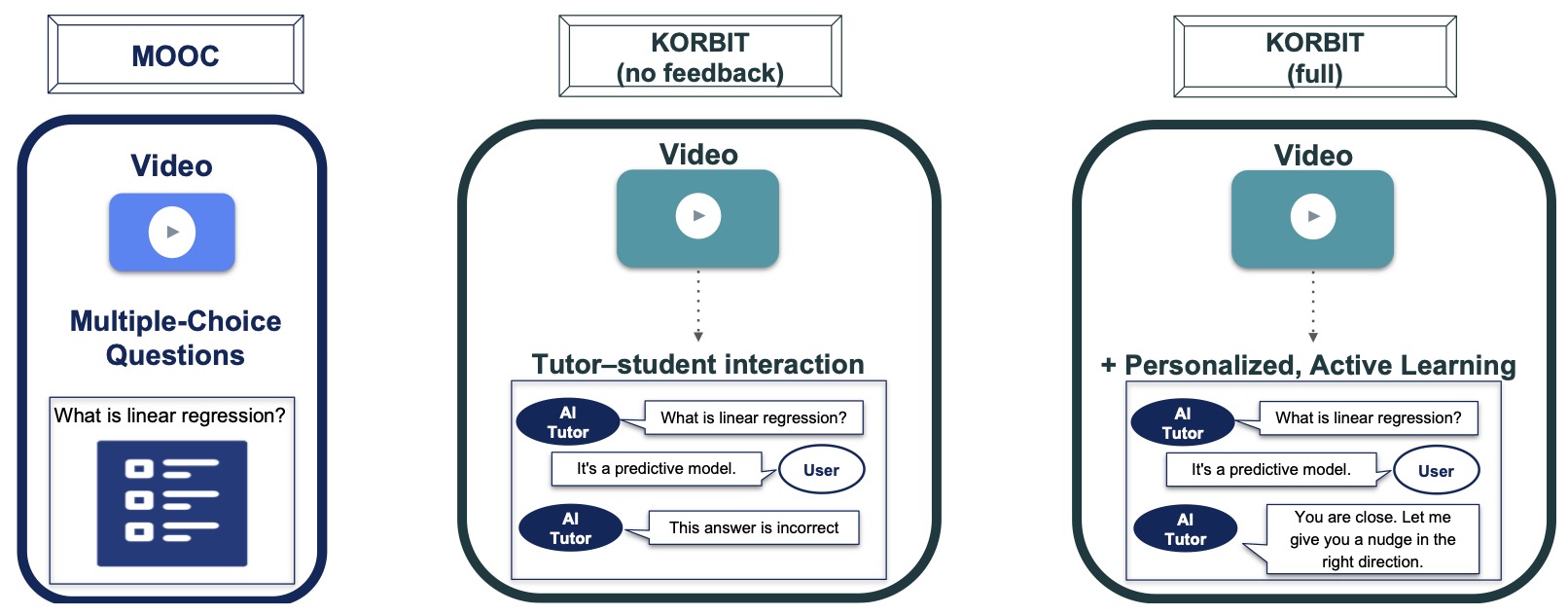}
\caption{A {\tt MOOC} platform follows a traditional learning approach, while {\tt Korbit (full)} uses a personalized, active learning approach with problem-solving and personalized feedback. Korbit also offers personalized Socratic tutoring, coding and project-based learning activities, though these learning activities were not considered in the present study. {\tt Korbit (no feedback)} is the version of the platform with the personalized feedback option disabled.} \label{fig:platforms}
\end{figure*}

Figure \ref{fig:platforms} visualizes the interaction on the platforms for the three groups of students.

\subsubsection{Selection of the Material}

One of the assumptions we made about the study process on the two platforms was that students benefit from personalized education. To test this assumption, we selected a set of participants, who had no or little prior experience with the topics and offered them to undertake a course on an introductory topic in data science. {\em Linear regression} was selected as the topic of study on both online platforms as such introductory topic: besides being one of the most fundamental topics in data science, it is typically covered early on in any course on the subject. To make sure the experiment provides for a fair comparison, we checked that the material covering this topic, as well as its difficulty level and the length of the courses on the two platforms are comparable and carefully aligned. Specifically, the courses on both platforms covered such sub-topics as numerical variables, correlation, residuals, least squares regression, and evaluation metrics, among other sub-topics.

Both the original course on linear regression on {\tt MOOC} and the adapted course on {\tt Korbit} consisted of short lecture videos on the subject, followed by multiple-choice questions in the case of {\tt MOOC}, and interactive problem-solving exercises in the case of {\tt Korbit}, either with a full interactive dialogue for {\tt Korbit (full)} or just the assessment of the answers for {\tt Korbit (no feedback)}
The course on each platform took approximately 3 hours to complete.



\subsubsection{Study Flow}

The study participants were recruited from the employees of the company in need of upskilling. Upon being selected for the study, they received an email with instructions about how to complete the study. They were asked to fill in the enrolment questonnaire to check their eligibility (only employees with little knowledge of data science were qualified to participate in the study). Then, eligible study participants were asked to complete a pre-assessment quiz on linear regression on TypeForm.\footnote{https://www.typeform.com} Once the participants were allocated to one of the study groups, they had one week to complete the $3$-hour course on the respective online platform. Upon completion of the course, they were asked to compete a post-assessment quiz on TypeForm. At the end of the study, all participants who completed it received an award (a completion certificate) regardless of their assessment scores.

Using pre- and post-quiz scores, we measured {\em learning gains} to quantify how efficiently each participant has learned. The pre- and the post-quizzes both consisted of $20$ multiple-choice questions, which were equally adapted to both courses. This was done in order to make sure that any topic mentioned in the quizzes was covered on both platforms to a similar extent, which means that participants that were using either of the platforms were expected to be able to successfully answer such questions.

We also ran a number of checks, and the quizzes passed a series of independent reviews to make sure that the questions were not inherently biased towards either of the two platforms in any way. Finally, questions of the pre-quiz were isomorphically paired with questions in the post-quiz to make sure that the difficulty of the two quizzes was as similar as possible without any questions being identical. This allowed us to measure the learning gains in a fair and unbiased way.

\subsubsection{Expected Outcomes}

We set out to test the following two hypotheses: 

\begin{blockquote}
{\bf Hypothesis 1}: 
{\tt Korbit (full)} results in higher learning outcomes than both {\tt MOOC} and {\tt Korbit (no feedback)}, which demonstrates that personalized feedback provides a more effective online learning experience.
\end{blockquote}

\begin{blockquote}
{\bf Hypothesis 2}: 
{\tt Korbit (no feedback)} results in higher learning outcomes than {\tt MOOC}, which demonstrates that problem-solving exercises provide a more effective online learning experience.
\end{blockquote}

\subsubsection{Learning Gains}
To evaluate which of the three learning setups teaches the participants more effectively, we compare {\tt MOOC}, {\tt Korbit (full)}, and {\tt Korbit (no feedback)} using {\em average learning gains} and {\em normalized learning gains}~\cite{hake1998interactive} of the participants under each setting. A student's learning gain $g$ is calculated as the difference between their score on the post-quiz and on the pre-quiz using the following estimate:

\begin{equation}
g = post\_score - pre\_score
\end{equation}

\noindent with $post\_score$ being the student's score on the post-quiz, and $pre\_score$ is their score on the pre-quiz. Both scores fall in the interval $[0\%, 100\%]$. A student’s individual normalized learning gain $g_{norm}$ is calculated by offsetting a particular student's learning gain against the score range in the ideal scenario in which a student achieves a score of $100\%$ in the post-quiz:

\begin{equation}
g_{norm} = \frac{post\_score - pre\_score}{100\% - pre\_score}
\end{equation}

\section{Results and Discussion}

In this section we summarize the results of our experiments, in particular by investigating the learning gains, study time and completion rates for each platform.
Higher learning gains show that the students benefit more from the learning process, while higher study time combined with a higher completion rate demonstrate higher levels of student engagement in the learning process.

\subsection{Student engagement}

\textbf{Study time} We first observe that participants on {\tt Korbit (full)} spend significantly more time studying and interacting with the platform. Specifically:

\begin{itemize}
    \setlength{\itemindent}{-1em}
    \item Study time {\tt Korbit (full)}: $141 \pm 28$ min.
    \item Study time {\tt Korbit (no feedback)}: $101 \pm 29$ min.
\end{itemize}

This result is statistically significant at $95\%$ confidence level ($p = 0.031$).

We note that the participants in these two groups use the same platform with respect to the study material (e.g., videos) and exercises. The core difference between these two versions of the platform is the amount of feedback that the participants get. In the case of {\tt Korbit (no feedback)}, the participants are only notified whether their answer to the question is correct or not, but are not provided with any further explanation. Note, that we do not have access to {\tt MOOC} for this metric and, therefore, cannot compare these results directly. However, since {\tt MOOC} does not provide students with active learning and personalized education, we would expect to see the results for {\tt MOOC} on a par with {\tt Korbit (no feedback)}. 

\subsubsection{Completion rates}

Next, we estimate completion rates for the material on each platform.
This is defined as the proportion of the number of participants who completed the full course with a skip rate below the threshold of $25\%$ to the number of participants who completed the pre-quiz and signed up on a platform for their group. We observe the following course completion rates:

\begin{itemize}
    \setlength{\itemindent}{-1.0em}
    \item Completion rate {\tt Korbit (full)}: $40.9\% \pm 11.9$
    \item Completion rate {\tt Korbit (no feedback)}: $29.4\% \pm  10.8$
    \item Completion rate {\tt MOOC}: $18.5\% \pm 9.4$
\end{itemize}

These results are significant with $p=0.082$ for {\tt Korbit (full)} ($n=66$) vs {\tt Korbit (no feedback)} ($n=68$), and $p=0.0025$ for {\tt Korbit (full)} ($n=66$) vs {\tt MOOC} ($n=65$). Combined with the results above, this demonstrates that the difference in the amount of feedback is the main reason for the higher completion rate. This is an important piece of evidence demonstrating the impact of feedback on the online study process.

\subsection{Learning Outcomes}

\begin{figure}[h!]
\centering
\includegraphics[width=0.5\textwidth]{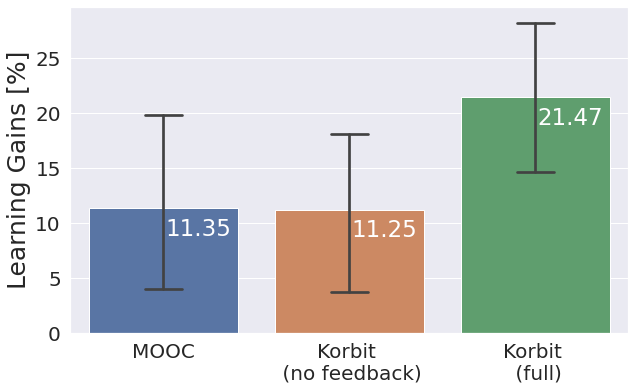}
\caption{Average learning gains $g$ with $95\%$ confidence intervals.} \label{fig:gains}
\end{figure}

\textbf{Raw Learning Gains}
We report average learning gains in Figure \ref{fig:gains} for the three study groups on the two learning platforms.\footnote{$95\%$ confidence intervals (C.I.) are estimated as:\\ 
\indent $\text{C.I.} = 1.96 * \frac{{Standard \ Deviation}}{\sqrt{{Population \ Size}}}$.} With respect to the raw learning gains $g$, we observe that both {\tt MOOC} and {\tt Korbit (no feedback)} show very similar learning gains of $11.25$–$11.35$. Both groups of participants, who used these platforms, were presented with exercises but not given feedback on their performance. At the same time, we observe that the learning gains on {\tt Korbit (full)}, which provided participants with personalized feedback and explanations relevant to their solutions, are around $90\%$ higher than on {\tt MOOC} with $95\%$ confidence ($p$=$0.04$), and around $92\%$ higher than on {\tt Korbit (no feedback)} with $95\%$ confidence ($p$=$0.046$).

\subsubsection{Normalized Learning Gains}

We observe a similar trend with respect to the normalized learning gains: the average normalized learning gains $g_{norm}$ for {\tt Korbit (full)} participants are $2.48$ times higher than the average normalized gains for {\tt MOOC} participants (with the difference being statistically significant at the $95\%$ confidence level ($p$=$0.01$). The key difference in these results compared to the raw learning gains is that {\tt Korbit (full)} contributes to higher learning gains than {\tt Korbit (no feedback)}, although the difference in this case is smaller: the improvement is $60.5\%$ with $p$=$0.15$.
 
\begin{figure}[h!]
\centering
\includegraphics[width=0.5\textwidth]{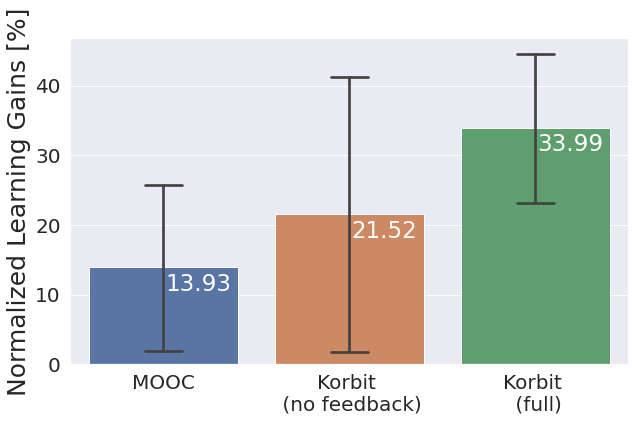}
\caption{Average normalized learning gains $g_{norm}$ with $95\%$ confidence intervals.} \label{fig:norm-gains}
\end{figure}

The results presented positively confirm both of our two hypotheses.
The results demonstrate that learning outcomes are better for participants on {\tt Korbit (full)} than participants on either of the platforms that do not provide personalized feedback.
This finding is confirmed by both the average learning gains $g$ and normalized learning gains $g_{norm}$, with these scores being substantially higher for {\tt Korbit (full)} than the other two platforms, with the difference in most cases being statistically significant at $95\%$ confidence level.
Further, the results presented demonstrate that learning outcomes are better for participants on {\tt Korbit (no feedback)} than participants on {\tt MOOC}, which shows the impact of active learning based on problem-solving exercises.
These results, combined with our earlier observations on increased study time, suggest that personalization and active learning elements on {\tt Korbit (full)} contribute significantly to learners' experience with online platforms. 





\section{Conclusion}


Artificial intelligence (AI) has contributed to the improvement of our society in many domains.
One domain which could and should benefit from the application of AI algorithms is education. Unfortunately, high-quality in-person teaching is not available to the majority of people around the world.
Even in the developed countries, it is currently not available to an increasingly large number of students due to the ongoing pandemic.
AI-powered online learning platforms can help bridge this gap.
This implies that AI researchers must focus their efforts on the application of AI to improve the quality of education, in particular in the context of online learning systems. ITS (intelligent tutoring systems) are a viable, scalable solution that can provide high-quality education, substantially surpassing the capabilities of platforms that do not allow for personalization and active learning.

In this paper, we have argued for the effectiveness of personalization and active learning strategies in ITS. Based on experimental evidence with nearly 200 students, we have shown that personalization and active learning in ITS can have a huge impact on both student learning outcomes and student motivation.
We have observed that learning gains are $2$ to $2.5$ times higher in an ITS compared to online learning platforms  without personalization and active learning.
We have observed that students are considerably more engaged in the learning process on the platform that provides highly personalized, active learning experience.
The latter is evidenced both by longer study time on the platform providing such personalized experience and by higher completion rates as compared to the alternatives that do not provide interactive, personalized education.

These results together provide strong evidence of the  tremendous impact that can be achieved with a personalized, active learning AI-powered system based on problem-solving exercises and personalized feedback.
If such a personalized and active learning experience can be made accessible to millions of learners around the world, then this advance would represent a huge leap forward towards the democratization of education in accordance with the United Nations' Sustainable Development Goals.


\bibliographystyle{named}
\bibliography{ijcai22}

\end{document}